\documentclass{paper}
\usepackage[utf8]{inputenc}
\usepackage[english]{babel}
\usepackage{graphicx}
\usepackage{amsmath}
\usepackage{listings}
\usepackage{mathtools}
\usepackage{quoting,xparse}

\NewDocumentCommand{\bywhom}{m}{% the Bourbaki trick
  {\nobreak\hfill\penalty50\hskip1em\null\nobreak
   \hfill\mbox{\normalfont(#1)}%
   \parfillskip=0pt \finalhyphendemerits=0 \par}%
}

\NewDocumentEnvironment{pquotation}{m}
  {\begin{quoting}[
     indentfirst=true,
     leftmargin=\parindent,
     rightmargin=\parindent]\itshape}
  {\bywhom{#1}\end{quoting}}

\graphicspath{ {support/} }
	
\title{Recommender Systems with Random Walks:\\ A Survey \footnote{This study has benefited from the guidance of Professor Wen Dong, Department of Computer Science and Engineering, SUNY at Buffalo.}}
\author{Laknath Semage \thanks {Department of Computer Science \& Engineering, SUNY at Buffalo.  Email: {\tt buddhika@buffalo.edu}}}
\date{\vspace{-5ex}}	
		
\begin{document}

\begin{titlepage}
	\maketitle	
	\thispagestyle{empty}

	\begin{abstract}

	Recommender engines have become an integral component in today's e-commerce systems. From recommending books in Amazon\footnote{https://www.amazon.com/} to finding friends in social networks such as Facebook\footnote{https://www.facebook.com/}, they have become omnipresent. 
	
	\vspace{2mm}	

	Generally, recommender systems can be classified into two main categories: content based and collaborative filtering based models. Both these models build relationships between users and items to provide recommendations. Content based systems achieve this task by utilizing features extracted from the context available, whereas collaborative systems use shared interests between user-item subsets. 
	
	\vspace{2mm}		
	
	There is another relatively unexplored approach for providing recommendations that utilizes a stochastic process named random walks. This study is a survey exploring use cases of random walks in recommender systems and an attempt at classifying them.
	
	\end{abstract}
	
\end{titlepage}	

	\pagebreak

	\section{Introduction}
	
	The general consensus on Recommender Systems (RSs) is that they are ``software tools and techniques providing suggestions for items to be of use to a user'' \cite{rec_handbook}. These systems are now ubiquitous in e-commerce portals and many other places where there is a relationship between users and items. The item of interest could vary from day-to-day items and commodities such as groceries, music and food to live entities such as friends. In the abstract sense, ``item'' signifies anything that is of interest to the user.

	\vspace{2mm}		
	
	The general idea behind recommender systems is to find the degree of similarity between users and items, and present some chosen \textit{n} most agreeable items to the user. Usually, this is done by representing characteristics of users and items as some selected set of features. Then, using mathematical distance calculation methods, the similarity between users and items is calculated to find matching items for a given user. 

	\vspace{2mm}		
	
	Based on the approach recommender systems take in finding associations between users and items, RSs are traditionally classified into two main groups.

	\vspace{2mm}
	
	\textbf{Content-based:} The system recommends items similar to those the user had an interest in the past. The interest could be expressed in many ways such as purchasing an item, clicking on the link of an item (e.g., in an e-commerce system) or the duration of time user has spent on the content of an item. For example, if a user has purchased some item with a certain characteristic, then the RS can suggest more items with the same characteristic as possible matches. 
	
	\vspace{2mm}

	\textbf{Collaborative filtering (CF):} These systems are based on the idea that a user with similar tastes to another user would have items that are of common interest to both users. The similarity between two users is calculated based on their history, usually using purchasing behaviors, click behaviors .etc. Then, user preferences are expressed as a selected set of features and matches are founds using people-to-people correlations \cite{ecommerce_systems}.
	
	\vspace{2mm}	
	
	There are other types of recommender systems, usually derivations or hybrids of content-based and CF based methods. Some of them are demographic based, knowledge-based, community-based and hybrid recommender systems \cite{rec_handbook}.
	
	\subsection{Random Walks}
		
	An ordinary random walk is a dynamic process run on a graph. The graph can be a $d$-dimensional integer lattice, and  time may be discrete or continuous. In the simplest case where $d=1$, the random walk can move a step to the right or the left at a given time with a defined probability $p$. Thus, the process produces a non-deterministic array of state changes \cite{rw_in_randome_env,rw_analysis}.
	
	\vspace{2mm}	
	
	The origin of random walks can be traced back to the Drunkard's walk problem \cite{drunkard_walk}. 
	
	\vspace{2mm}	
	
	\begin{pquotation}{Karl Pearson, 1905}
``The lesson of Lord Rayleigh's solution is that in open country the most probable place to find a drunken man who is at all capable of keeping on his feet is somewhere near his starting point!

	\vspace{2mm}	

	A man starts from a point $O$ and walks $l$ yards in a straight line; he then turns through any angle whatever and walks another $l$ yards in a second straight line. He repeats this process $n$ times. I require the probability that after these $n$ stretches he is at distance between $\tau$ and $\tau+\delta\tau$ from his starting point $O$'' \cite{drunkard_walk_nature}
	
	\end{pquotation}		
	
	\vspace{2mm}	

	Presently, the problem is usually expressed as the sum of a sequence of independent and identically distributed (IID) random variables. 

	\vspace{2mm}		
	
	Let $\{X_i; i \geq 1\}$ be a sequence of IID random variables, and let $S_n = X_1 + X_2 + ... + X_n$. The integer-time stochastic process $\{S_n; n \geq 1\}$ is called a random walk, or more precisely, a one-dimensional random walk based on $\{X_i; i \geq 1\}$ \cite{stochastic_book}.

	\vspace{2mm}		
	
	Though $S_n$ is a simple random variable for any $\{S_n; n \geq 1\}$, the focus is not only on $S_n$, but also on the behavior of the entire random walk process. For example, for a given real number $\alpha$, the probability that a sequence will contain any term for which $S_n \geq \alpha$, or in other words, the point of threshold crossing is a common question that can be answered through a random walk process.

	\vspace{2mm}			
	
	As a quick example, let's consider a simple random walk: if $X_i$ are  IID binary random variables with values 1 or -1, the probability $S_n$ will take some value can be answered as follows. If $m$ out of $n$ trials are 1, then $S_n = m - (n - m); S_n = 2m - n$. If $p$ is the probability of $X_i=1$, then,
	\begin{equation*}
		Pr(S_n = 2m - n) = \frac{n!}{m!(n-m)!}p^m(1-p)^{n-m}
	\end{equation*}
	
	The same concept can be extended to integer-valued random walks, where $X_i$ are IID integer values instead of binary values \cite{stochastic_book}.	
	
	\vspace{2mm}		

	\section{Random Walks in Recommender Systems}

	In the literature of recommender systems, random walks have been predominantly used in the context of markov models or graphs. The general idea has been to represent relevant entities and relationships of a given problem as a graph structure and infer additional relationships to provide useful recommendations. Random walks are particularly suited for this purpose due to some of the properties they hold which will be discussed later in this paper.
	
	\vspace{2mm}	
	
	Based on the usage of random walks in recommender systems, we can identify several key categories.
	
	\begin{enumerate}
		\item Global ranking methods
		\item Random walks with restarts (RWR)
		\item Absorbing random walks
	\end{enumerate}

	\subsection{Global Ranking Systems}
	
	Global ranking methods provide a single ranking of elements for an entire system. Thus, they are not customizable for individual preferences. PageRank \cite{pagerank}, the widely used search ranking algorithm is an example for such a global ranking method, and can be expressed as a formulation of a global random walk.
		\vspace{2mm}		

	The PageRank algorithm can be intuitively thought of as a random surfer. This random surfer would keep following links in a page depending on the importance of links. Also there is a relatively small probability that the surfer would jump to some other page not pointed by a link. Thus, the PageRank random walk is a stochastic process where at any given time it could be at a particular node and in the next time step, with a defined probability, choose an outedge uniformly at random to traverse to the next node. Then we are interested in the convergence of this process to a limiting distribution on the set of nodes in the graph. Using the stationary distribution of this process, the importance of nodes in the graph can be ranked, in a manner that's common to all users.
		\vspace{2mm}		

	In the formulation of the graph $G = (V,\xi)$, a link from a web page $u \in Web$ to $v \in Web$ ($u \rightarrow v$) implies $v$ is an important page for $u$. The level of importance is proportional to the importance of $u$, and inversely proportional to the number of outgoing links from $u$, which can be denoted as $deg(u)$. If the surfer is at $u$ at $k$ time step, the surfer lands at node $v_i \in \{v|u \rightarrow v\}$ at time stamp $k+1$ with probability $1/deg(u)$. Under this formulation, the PageRank of page $i$ is defined as the probability that at any given time step $k$, the surfer being at page $i$. Considering the markov chain induced by this random walk, the transition probability from page $i$ to some other page $j$ can be given as $P_{ij} = 1/deg(i)$ \cite{pagerank2}.
		
		\vspace{2mm}		
		
	With the above definition for the transition probability, for $P$ to be a valid transition matrix, each node must have at least one outgoing link. In other words, there must not be any nodes with 0 outdegree, or commonly known as sink nodes. However, this property usually does not hold true for web graphs since there can be web pages that don't have any outgoing links. A transition matrix of this nature can be converted to a valid transition matrix by adding a uniform transition probability mass for all sink nodes \cite{pagerank2}.
	
		\vspace{2mm}		

By the Ergodic theorem for Markov chains, a Markov chain has a unique stationary probability distribution if it's aperiodic and irreducible \cite{pagerank2}. In a regular web graph, the property of irreducibility generally does not hold true because in most cases web graphs are not fully connected. By adding links from all nodes to every other node with a small transition probability, the PageRank algorithm transforms a web graph into a strongly connected graph, which supports aperiodic and irreducible properties \cite{web_data_mining}.

		\vspace{2mm}		

	According to the above definition, the PageRank score for some node $n$ can be defined as,
	
	\begin{equation}
		PR(n) = \alpha.\sum_{q:(q,n)\in\xi} \frac{PR(q)}{deg(q)} + (1-\alpha).\frac{1}{|V|}
	\end{equation}
	where $\alpha$ is called the decaying factor, a constant for which 0.85 is a common choice \cite{itemrank}. The term $(1 - \alpha)/|V|$ signifies the transition probability from any node in the graph to the node $n$ (\textit{teleportation}). 
			
		\vspace{2mm}		

	Since global ranking methods such as PageRank provide only a single ranking for an entire system, they can not be used for personalizing tasks. However, such methods are an important tool when nodes in a graph need to be ranked according to their importance. In turn, such rankings can be used for recommendations \cite{pagerank_rec_class}.
	
	\subsection{Random Walks With Restarts}
	
	The general PageRank method discussed in 2.1 does not consider personal preferences, nor it customizes for individual tastes. However, recommender systems are generally expected to take personal preferences into consideration. \textit{Random walks with restarts (RWR)} provides random walks the capability of personalization. 
	
	\vspace{2mm}		
	
	In regular random walks over a graph, the random walker jumps across nodes based on the graph's transition probability matrix. Given that the linked structure is irreducible and aperiodic, the random walk will converge to a unique stationary distribution. Therefore, the resulting steady-state probabilities are a function of both the graph structure and transition probabilities. This is the basis of global ranking methods such as PageRank; Random walks with restarts is an extension of general random walks taking user preferences into consideration. It achieves this by simply adding a constant probability of jumping back to the originating node at each time step. This gives the random walk the behavior that nodes closer to the originating node are ranked higher than distant nodes, thereby giving the walker a personalized view of the network \cite{rwr_city}.

	\vspace{2mm}		
	
	The PageRank equation of (1) can be generalized as:

	\begin{equation}
		Q = \alpha.W + (1-\alpha)R
	\end{equation}	
	
	$W$ encodes the structure of the graph and represents transition probabilities of nodes, and $R$ encodes the randomness of the walk where the walker can randomly jump to any other node. While in regular random walks $R$ takes a common value for the whole graph, in biased random walks such as random walks with restarts, $R$ encodes the preference for the jump \cite{personlized_pr}. 
		
	\vspace{2mm}

	The formulation of RWR for recommendations can be made using directed or undirected graphs. For example, \textit{A. Noulas et al.}\cite{rwr_city} used RWR to recommend venues to visit using location based social networks and geographical data. For this task they modeled data as an undirected graph, with users and venues as nodes. If a user had visited a particular venue it had a non-zero edge, and two users were linked if they were friends. Then this graph was used to define transition probabilities uniformly for each edge. Then for every user $i$ they defined a random walk with restarts such that at every step there was a constant probability of jumping back to the node representing user $i$. In order to generate recommendations for the user $i$, they had computed steady-state probabilities for the user, and ranked venues in decreasing order. This process gave priority to locations that were highly connected through friends, checked-in venues or a combination thereof. Moreover, they had also attempted a weighted version of the random walk, where transition probabilities of edges going out from a node were not uniform. However, using a similar graph setup \textit{M. Ye et al.} \cite{rwr_poi} compared RWR with several other collaborative filtering methods and reported that RWR may not be suitable for point of interest (POI) recommendations in location based social networks (LBSNs), because of significantly differing preferences of users in their social network graph.
	
	\vspace{2mm}		
	
	Another application of RWR is Folksonomy based recommendations by \textit{I. Cantador et al.}\cite{rwr_folksonomy}. In this they represented users, items and tags as nodes in the graph. The relationships between nodes were encoded using uni-directional or bi-directional edges. Then a random walk with restarts was executed to get stationary probabilities and rank items according to the predicted preference. A similar approach was taken by \textit{I. Konstas et al.}\cite{rwr_cf} on music track recommendations by representing users, tracks and tags as nodes in the graph. In this study, they encoded user's tagging behavior, social network relationships and music tastes using edges in the graph. For example, relationships between users were represented as bi-directional edges indicating a mutual friendship, and music tracks a user listened to were represented by edges weighted by the number of times the track was listened.

	\vspace{2mm}		
	
	There are many similar applications of RWR because of the advantages it provides. Similar to the PageRank algorithm, RWR based algorithms such as personalized PageRank \cite{personlized_pr} also maintain propagation and attenuation properties. Therefore, they are ideal to represent the decaying nature of preference as the distance increases between a given user node and target nodes. Since they can find a measure of relevance even when two nodes are not directly connected, random walks with restarts provide a solution for the sparsity problem which is very common in recommender systems. Specially when the data contains a large number of dimensions, the sparsity issue can be a major challenge and RWR provide a way to handle such data. Another advantage of RWR is the capability to blend influences of various factors. Therefore, it has become a popular method when it's required to combine various factors such as social network relationships and user preferences. Furthermore, rankings can be pre-computed using RWR so it can be scaled to large data sets \cite{rwr_multidim}.

	\vspace{2mm}		

	Going beyond the original concept, random walks with restarts have been used in many instances with various extensions or modifications \cite{rwr_multidim,ecommerce_rwr,trust_based_rw}.
	
	\subsubsection{Support for Multi-dimensional Queries}
	
	The personalized PageRank algorithm \cite{personlized_pr} is only capable of providing a ranking of entities for targeted nodes which are usually users in the system. Extending the graph formulation and introducing a query vector to the personalized PageRank, \textit{Lee et al.}\cite{rwr_multidim} have adapted it to rank entities according to a given query. 

	\vspace{2mm}		

	The problem formulation is as follows: instead of the usual $User$ x $Items$ dimensional space, the entire domain space $\varepsilon$ has been considered as a union of disjoint domains $D_i$: $D_1 \cup D_2 \cup ... \cup D_n$. Then entities in the domain $D_i$ are represented as $e_{i1},e_{i2},...,e_{ik}$. A query $Q$ is defined as a subset of $\varepsilon$ and the measure of relevance score between $Q$ and entity $e_{ij}$ is represented as $u(Q,e_{ij})$ where $u$ is a utility function. Then to recommend entities for the query, top-k entities that maximize $u(Q,e_{ij})$; $e_{ij} \in D_T$ can be selected where $D_T$ is the target domain. For example, if the dataset has domains (attributes) \textit{USER, LOCATION, WEATHER, SONG}, and we want to recommend \textit{SONG} to the $u \in USER$ factoring in current location $l \in LOCATION$, current weather $w \in WEATHER$, the query can be defined as $Q = \{u,l,w\}$ and the target domain as \textit{SONG}.
	
	\vspace{2mm}		
	
	To encode these information, authors have constructed a bipartite graph where one disjoint set represents nodes from the target domain and the other represents nodes from the non-target domain (recommendation factors) that participate in the recommendation process. Then edges have been constructed between nodes in the target domain and recommendation factor nodes using co-occurrences of entities, and normalized to construct the transition matrix $P$.
	
	\vspace{2mm}		

	To rank entities a modified random walk with restart has been used. The PageRank algorithm from (1) can be written using vectors as,
	
	\begin{equation}
		\vec{r} = \alpha.P^T\vec{r} + (1-\alpha).\frac{1}{n}\vec{e}
	\end{equation}
	where $P$ is the transition matrix, $\vec{r}$ is PageRanks of nodes, $\alpha$ is the damping factor, $n$ is the number of nodes and $\vec{e} = \{1,1,..,1\}^T$. 
	
	\vspace{2mm}		
	
	For personalized PageRank, $\frac{1}{n}\vec{e}$ in (3) is substituted with personal teleport $\vec{t}$ representing that a random walker can teleport to the node in interest instead of a random node in the graph. Thus, the Personal PageRank can be written as,

	\begin{equation}
		\vec{r}_u = \alpha.P^T\vec{r} + (1-\alpha).\vec{t}
	\end{equation}
	where $t_i$ is 1 if the node is user's interest, and 0 otherwise.

	\vspace{2mm}			
	
	Following the same principle, this study has replaced personal bias $\vec{t}$ with a normalized query vector $\vec{q}$ to encode domains interested in the query. Unlike $\vec{t}$, however, $\vec{q}$ can take more than one 1s representing multiple domains interested. Thus, the modified page rank follows the form,
	
	\begin{equation}
	\begin{split}
		\vec{r} = \alpha.P^T\vec{r} + (1-\alpha).\vec{q} \\
		\tilde{q_i} = \begin{cases}
		\begin{aligned}[c]
			1, \quad if \varepsilon_{v_i} \in Q \\
			0, \quad otherwise
        	\end{aligned}
     	\end{cases}
     \end{split}
	\end{equation}
	where $\tilde{q}$ is the unnormalized $\vec{q}$, and $\varepsilon_{v_i}$ is the set of entities that node $v_i$ associates with.
	
	\vspace{2mm}			
	
	This process gives a higher ranking to a node if it's closely related to nodes in the query. Therefore, by ordering $r_t$ for entities $e_t \in D_T$, it's possible to find top-k terms relevant to the query.
	
	\subsubsection{Supervised Random Walks}
	
	One of the advantages of random walks with restarts is the ability to combine graph data such as social network information with the rest of data. \textit{Backstrom et al.} \cite{rwr_supervised} have presented a way to combine RWRs with supervised learning to predict nodes, to which a user is likely to make new links in the future. They have done this by learning a function that assigns strengths to edges in the network such that when the random walk is run, the walker is more likely to visit nodes to which new links will be added in the future (i.e., positive nodes).  

	\vspace{2mm}		

	Formally the problem and the approach are as follows:

	\vspace{2mm}		
	
	Given a directed graph $G(V,E)$, a source node $s$ and a set of candidate nodes $C$, the goal is to identify the set of nodes $D = \{d_1,...,d_k\}$ to which the source node $s$ will create edges in the future. Moreover, if the set of nodes to which edges are not created (no-link nodes) is $L = \{l_1,...,l_2\}$, then $C = D \cup L$. Each edge $(u,v)$ in $G$ has a corresponding feature vector $\psi_{uv}$ that describes the nodes $u$ and $v$ (e.g., age, gender, hometown) and/or interactions between nodes (e.g., number of messages passed between two nodes).
	
	\vspace{2mm}		

	For edge $(u,v)$, the framework introduces edge strength $a_{uv} = f_w(\psi_{uv})$, where the function $f_w$ parameterized by $w$ takes the edge feature vector $\psi_{uv}$ as input. The random walk with restart will then use these edge strengths as transition probabilities in its run from $s$. The stationary distribution $p$ of the random walk assigns each node $u$ a probability $p_u$ which then used for ranking. The objective of supervised learning is to learn parameters $w$ such that, $d \in D, l \in L: p_l < p_d$.
	
	\vspace{2mm}		
	
	\begin{equation}	
	\begin{aligned}
		\min_w F(w) = ||w||^2 + \lambda \sum_{d \in D, l \in L} h(p_l - p_d)  \\
		\textrm{such that}\\
		\forall d \in D, l \in L: p_l < p_d	\\
	\end{aligned}		
	\end{equation}

	where $p$ is the vector of PageRank. The shortest $w$ parameter vector is considered for the purpose of regularization. $h$ is a loss function of choosing that penalizes violated constraints. It assigns a non-negative penalty considering the difference $p_l - p_d$: if $p_l - p_d < 0$ then $h(.) = 0$, while for $p_l - p_d > 0, h(.) > 0$. $\lambda$ is a regularization parameter that trades-off between the complexity and the fit of the model. 
	
	\vspace{2mm}	
	
	Edge strengths can be transformed to stochastic transition matrix $\acute{Q}$:
	
	\begin{equation}
		\acute{Q}_{uv} = \begin{cases}
				\begin{aligned}[c]
               		\frac{a_{uv}}{\sum_w a_{uw}}  \\
                		0
                \end{aligned}
                \qquad\
                	\begin{aligned}[c]
                		if (u,v) \in E, \\
                		otherwise
                \end{aligned}

            \end{cases}
	\end{equation}	

	By incorporating restart probability $\alpha$, the final random walk transition probability matrix $Q$ can be obtained. $Q$ defines the conditional probability that a walk will traverse the edge $(u,v)$ given it's currently at node $u$.

	\begin{equation}	
		Q_{uv} = (1 - \alpha) \acute{Q}_{uv} + \alpha 1(v = s)
	\end{equation}	
	
	The vector $p$ is the stationary distribution of the random walk, and is the solution to following eigenvector equation:
	
	\begin{equation}
		p^T = p^T Q
	\end{equation}	

	To minimized eq. 6 (optimization problem), gradient of $F(w)$ is derived with respect to $w$, and gradient based optimization methods can be used to solve the problem.
	
	\begin{equation}
	\begin{aligned}	
		\text{Let}\ \delta_{ld} = p_l - p_d \\
		\frac{\partial F(w)}{\partial w} = 2w + \sum_{l,d} \frac{\partial h(\delta_{ld})}{\partial \delta_{ld}} (\frac{\partial p_l}{\partial w} - \frac{\partial p_d}{\partial w} ) \\
	\end{aligned}
	\end{equation}
	
	To find partial derivatives of $\frac{\partial p_d}{\partial w}$, from Eq. 9,
	\begin{equation}
	\begin{aligned}
		p_u = \sum_j p_jQ_{ju} \\
		\frac{\partial p_u}{\partial w} = \sum_j Q_{ju}\frac{\partial p_j}{\partial w} + p_j\frac{\partial Q_{ju}}{\partial w}	
	\end{aligned}
	\end{equation}
	
	To obtain $\frac{\partial Q_{ju}}{\partial w}$, from Eq. 7 and 8, when $(j,u) \in E$,
	\begin{equation}
		\frac{\partial Q_{ju}}{\partial w} = (1 - \alpha) \frac{\frac{\partial f_w(\psi_{ju}) }{\partial w} ( \sum_k f_w (\psi_{jk}) ) - f_w (\psi_{ju}) (\sum_k \frac{\partial f_w(\psi_{jk}) }{\partial w} ) }{ ( \sum_k f_w (\psi_{jk}) ) ^ 2  }
	\end{equation}
	and otherwise $\frac{\partial Q_{ju}}{\partial w} = 0$.
	
	By recursively applying the chain rule (11 and 12), $\frac{\partial p_u}{\partial w}$ can be iteratively computed by converging to a defined threshold. 

	\vspace{2mm}	
	
	By applying gradient descent methods $F(w)$ can be minimized. Since the problem is not convex in general, it's not guaranteed to find the global minimum and in practice will need multiple random initializations to find a good solution.
	
	\subsubsection{Decentralized Random Walks}
	
	Decentralized recommender systems offer solutions to several key challenges that hamper centralized recommender systems. Firstly, resource requirements for a centralized system can be avoided by distributing the computation of predictions among users. Secondly, a distributed recommender system can improve the privacy of its users due to the lack of a centralized system where a single breach of security can compromise all information. 

	\vspace{2mm}	
	
	However, applying random walks in a distributed setting has its own complications compared to a centralized system. Since each peer participating in the calculation has access to only a limited amount of information about its neighborhood, regular collaborative filtering (CF) algorithms can't be applied as they are applied in a centralized setting.  

	\vspace{2mm}	

	\textit{A.Kermarrec et al.}\cite{rwr_decentralized} have proposed a method of using epidemic protocols (gossip protocols) to estimate unseen neighbors, which has made it possible to execute CF algorithms in a decentralized setting. The epidemic protocol's behavior is as follows: Each peer in the system maintains details (a view) about a limited number of peers that maximize a clustering function. In the context of recommender systems, this can be a similarity measure such as \textit{Cosine similarity} or \textit{Pearson correlation} applied on the rating behaviors. At the joining, new peer's view is randomly initialized. Thereafter, the peer periodically selects peers from her view, exchanges view information and compares them with peer's currently held view information in order to improve the quality of her view. In addition, the same information exchange and comparison process is carried out against a set of randomly selected peers, and the peer's view is updated keeping $c$ most similar entries. This process is run until it converges.
		
	\vspace{2mm}	
	
  	For a user to compute a personalized score for an item, she employs scores assigned by users in her view and users connected within a defined number of hops (neighborhood). By running a random walk process on the Markov chain induced by the graph (considering users as vertices), the topology of users can be combined with traditional similarity measures such as \textit{Cosine similarity} to calculate the final similarity score between users. Unlike a centralized setup where all users' preferences need to be calculated, here only preferences of the neighborhood are considered, which in turn makes it possible to work with large datasets. Authors have observed that increasing the size of the neighborhood improves the precision of recommendations.
  	
	\vspace{2mm}	

	For the random walk, transition probability matrix ($P^a$) for user $a$ is calculated as follows:

		\[
		p_{uv}^a = (1 - \beta) \frac{\acute{s}_{uv}}{ \sum_{z\in K(a)} \acute{s}_{uz}} + \frac{\beta}{m}, \quad		
		\acute{s}_{uv} = \begin{cases*}
               	s_{uv}  & if $s_{uv} \geq 0$ and $u \neq v$ \\
               	\gamma_u & if $u = v$ \\
                	0 & otherwise
     	\end{cases*} \]%	,
     	
     	Where $K(a)$ is all users in the neighborhood of $a$, $s_{uv}$ is the similarity between $u$ and $v$, $\beta \in (0, 1)$ is the probability that user randomly jumps to any peer in the neighborhood, $m$ is the neighborhood size, and $\gamma$ represents the parameter where the same user answers the recommendation query.

	\vspace{2mm}	

	In a Markov chain, the probability of being in a given state depends only on its previous state. Therefore, if each peer decides to continue the random walk with a probability of $\alpha$, the initial transition probability of node $a$ over its neighborhood is $\vec{d}$ and $P^k$ is the power $k$ of transition probability matrix,
	
	\[
		Pr(X_k = u) = \alpha \sum_{v=1}^m Pr(X_{k-1} = v)p_{vu} = \alpha^k \sum_{v=1}^m \vec{d} P_{vu}^k
   	\]%   	

	When $k = \infty $ (long enough iterations), this method approximates the final probability distribution over the neighborhood. 
	
	\vspace{2mm}	
	
	Authors have observed that the random walk method discussed here yields comparably better results than classic similarity measures when the training set proportion is below 70\% (higher sparsity) with a view size of 30. When it comes to the view size, they have identified 30 as the best performing size.

	\subsection{Absorbing Random Walks}
	
	A traditional random walk without constraints can be used to accomplish tasks such as global ranking of all nodes in a graph. Such random walks, however, will produce the same statistics regardless of the starting point, and thus, are not suited for personalization tasks. Furthermore, if the goal is only to rank items preferred by users, a typical random walk lacks the ability to focus on a particular set of nodes in the graph (e.g, items). To address these shortcomings, \textit{A.P. Singh et al.}\cite{absorbing_rw} have suggested using random walks with absorbing states. In this study, they have utilized both social graphs (e.g., friends) of gamers and the fact of game's ownership for the task of providing recommendations for online games.
	
	\vspace{2mm}

	According to the proposed method, users and items, which in this case games owned by users, are represented as nodes in an augmented bipartite graph, where users form one side of the graph and items the other. Unlike a regular bipartite graph, however, edges connect users-to-users representing their social connections, and user-to-items representing the ownership of games. Formally, given sets $U = \{u_1, ..., u_n\}$ and $I = \{g_1, ..., g_n\}$ of users and items, the ownership graph $G = (U,I,O,S)$ where,
	
		\[
			O = \{(u,g): (u,g) \in U \times I\} \\			
		\]%
		\[
			S = \{(u,u'): (u,u') \in U \times U\}	
		\]%
	
	\vspace{2mm}	
	
	The induced graph is then transformed to a Markov chain with absorbing states in the following way. The random walk will start from the user for whom recommendations need to be produced; with $\alpha$ probability transitioning to a friend, or with $(1-\alpha)$ probability transitioning to an item. If the walker chooses to transition to a friend node, one is chosen uniformly at random from the set of friend nodes current user node is connected to. Similarly, if the walker decides to transition to an item, the item will be chosen uniformly at random from the list of items user is connected to (games user owns). Items are considered absorbing states: once the walker transitions to an item state, it can not leave the state. Formally, the absorbing random walk can be defined as:	
	
	\begin{equation}	
	\begin{aligned}		
		P(u'|u) =& \begin{cases*}
               	\frac{\alpha}{|S(u)|} & if $u' \in$ S(u) \\
                	0 & otherwise
     	\end{cases*}  \\
		P(g|u) =& \begin{cases*}
               	\frac{(1-\alpha)}{|O(g)|} & if g $\in$ O(u) \\
                	0 & otherwise
     	\end{cases*}  \\    	
		P(g'|g) =& \begin{cases*}
               	1 & if $g'$ = $g$ \\
                	0 & otherwise
     	\end{cases*}  \\  	
		P(u|g) =& 0 \\
	\end{aligned}			
	\end{equation}
	
	\vspace{2mm}	

	For a walk starting with user $u$ and converges to an absorbing distribution $P_u(g)$ over items $g$, the ranking problem for the user $u$ can be solved by ranking items in the decreasing order of $P_u(g)$ probabilities. From this list of rankings, once items owned by the user are removed, a valid ranking of recommendations for the user $u$ can be obtained.
	
	\vspace{2mm}	

	Utilizing the property of starting node critically influencing the ending absorbing state, \textit{S. Baluja et al.}\cite{youtube_absorbing_rw} have proposed a random walk based method of labeling videos in YouTube\footnote{http://www.youtube.com/}. They have used a video-video co-view graph where videos are vertices and edges connect videos commonly co-watched cutoff by a defined threshold. A subset of nodes (videos) in this graph are associated with labels, and the goal is to produce the most likely label distribution for unlabeled nodes. 
	
	\vspace{2mm}	
	
	The above graph $G$ can be formally represented as $G = (V, E, w)$ where $V$ denotes all nodes, $E$ all edges and $w$ a non negative weight on edges. Let $L$ be the set of labels and $V_L \subset V$ where $V_L$ is the set of nodes with labels. For the purpose of executing a random walk, for each $v \in V_L$ a ``shadow'' vertex $\tilde{v}$ is introduced and connected with edge $(v, \tilde{v}$) with a weight of 1. Then the label distribution $L_v$ for each node $v \in V_L$ is removed from the node $v$ and moved to $\tilde{v}$.

	\vspace{2mm}	
	
	These shadow vertices are considered absorbing states of Markov chain defined by the random walk. The labeling algorithm starts a random walk from each unlabeled vertex $v$ and outputs a labeled distribution $L_v$ upon reaching an absorbing state. Since the label distribution for each $v$ is a random variable, the final label distribution is yielded from the expectation of this random variable after running the random walk multiple times.

	\subsubsection{Biased Absorbing Random Walks}
	
	The absorbing random walk approach discussed so far considers user-user transition $P(u'|u)$ (13) as uniform at random, representing all social links carry the same weight or influence when making recommendations. However, this is usually not the case, and does not reflect the reality. To account for this fact, instead of transitioning to linked users uniformly at random, the choice can be biased by the similarity between the current user and the destination user. Therefore, the user-user transition probability $P(u'|u)$ in (13) can be replaced by the following:

	\begin{equation}	
	\begin{aligned}		
		P(u'|u) =& \begin{cases*}
               	\alpha \frac{e^{\beta(cos(g(u), g(u')) } }{ \sum_{u'' \in S(u)} e^{\beta(cos(g(u), g(u'')) } } & if $u'$ $\in$ S(u) \\
                	0 & otherwise
     	\end{cases*}  \\
     	where \\
     	cos(g(u), g(u')) =& \frac{g(u).g(u')}{|g(u)||g(u')|} 
	\end{aligned}	
	\end{equation}
	
	$g(u)$ is the binary vector representing items owned by the user $u$, and $\beta$ is a parameter that represents the likelihood of transitioning to a friend with similar tastes in games. Moreover, according to the above definition, unbiased absorbing random walk is a special case of biased absorbing random walk where $\beta$ = 0 \cite{youtube_absorbing_rw}.

	\section{Discussion}
	
	  One common characteristic of random walks shared among use cases discussed throughout this study is the controlling of preference flow between entities in the graph. For this, algorithms usually possess two key attributes: $propagation$ and $attenuation$ \cite{itemrank}. Propagation signifies the flow of positive and negative influences through the graph. However, in most applications the influence should decrease with the distance between nodes which signifies an attenuation.

	\vspace{2mm}	

	Random walks have been generally used in recommender systems when there is a need to elicit associations out of a graph structure (e.g., social network). By observing their behavior, random walks can be classified into two main categories: non-personalized and personalized random walks. The widely used web search ranking algorithm PageRank \cite{pagerank} is an example for non-personalized random walks. In the context of making recommendations, they can produce a global ranking of elements involved and provide recommendations in a non-personalized way. 
	
	\vspace{2mm}	
	
	Personalized random walks are predominantly based on the concept of random walks with restarts (RWR), where instead of a uniform teleportation probability distribution over all nodes, a biased probability is used on the starting node. This modification gives random walker the behavior of surfing close to the starting node, making it possible to personalize for individual users. Derivatives of random walks with restarts have been studied for answering multi-dimensional queries, link-predictions with supervised machine learning and decentralized personalized recommenders. In addition to RWRs, another modification introducing absorbing states to regular random walks has given rise to absorbing random walks. They have been also used in making personalized recommendations, but with an advantage over RWRs in that they can control the set of items over which recommendations are made.
	
	\vspace{2mm}	

	As a whole, one of the key advantages of random walks is the intuitive and explainable solution they present. They also provide a convenient method to combine graph and non-graph data for making recommendations. Further, studies have shown that some random walk methods perform better than classic similarity measures when the sparsity is higher in the dataset \cite{rwr_decentralized}, which is a common issue with recommender systems.
	
	\section{Conclusion}
	
	This paper is an attempt at surveying the domain overlapping recommender systems and random walks. Though it may be a relatively less studied domain compared to topics such as matrix factorization in the context of recommender systems, there is a wealth of literature on applying random walks for recommendations, especially when a graph structure is involved in the problem domain. During this study, we have identified three key categories based on the type of random walk used: global random walks, random walks with restarts and absorbing random walks. Each of the categories has its own specific use cases and we have explored them in depth, identifying derivatives and how they differ from each other. This study has not delved deep into performance concerns or comparisons against other non-random walk methods, and it remains an open problem for a future study.
	
\bibliographystyle{IEEEtran}	
\bibliography{References}

\iffalse

\fi

\end{document}